\documentclass[runningheads]{svmult}

\usepackage{makeidx}   
\usepackage{graphicx}  
\usepackage{subeqnar}  
\usepackage{multicol}  
\usepackage{physprbb}  
\makeindex             

\begin{document}
\title*{High energy photon flux prediction 
from neutralino 
annihilation in the globular cluster Palomar 13}
%
%
%
%
\titlerunning{Neutralino prediction in Palomar 13}
%
\author{Edmond Giraud\inst{1}
\and George Meylan\inst{2}
\and Mariusz Sapinski\inst{1}
\and Alain Falvard\inst{1}
\and Agnieszka Jacholkowska\inst{1}
\and Karsten Jedamzik\inst{3}
\and Julien Lavalle\inst{1}
\and Eric Nuss\inst{1}
\and Gilbert Moultaka\inst{3}
\and Frederic Piron\inst{1}
\and Pierre Salati\inst{4}
\and Richard Taillet\inst{4}
}
\authorrunning{Giraud, Meylan, Sapinski et al.}
%
%
\institute{GAM, Univ. Montpellier II, Place E. Bataillon, 34 095
Montpellier Cedex, France
\and STScI, 3700 San Martin Drive, Baltimore MD 21218, USA
\and LPMT, Univ. Montpellier II, Place E. Bataillon, 34 095
Montpellier Cedex, France 
\and LAPTH, Chemin de Bellevue, BP 110, Annecy-le-Vieux Cedex, France 
}

\maketitle              

\begin{abstract}
The distant globular cluster Palomar 13 has been found to have a very high 
mass-to-light ratio and its profile can be well fitted either by a King model 
with a tail, or with a NFW model [1]. This cluster may be the first case of 
the many clumps predicted by CDM simulations that would not be disrupted by 
the galactic halo potential. We make the hypothesis that Pal 13 is made of 
neutralinos and run the DarkSuspect code to estimate the high-energy photon 
flux due to the annihilation of neutralinos through various channels in some 
benchmark scenarios. These low fluxes may be used as targets to be reached 
in proposals for future ground-based high altitude Cerenkov telescopes.

\end{abstract}

\section{Introduction}
\vspace*{-0.1cm}
The distant globular cluster Palomar 13 has been found to have a very high
M/L ratio of $40~M_\odot/L_\odot$ and its profile can be well fitted either
by a King profile with a power-law tail or a NFW model [1] with scale radius 
$2.4 \pm 0.2$pc and central density $80~ M_\odot pc^{-3}$. A 
possible explanation is that this distant cluster $(D = 24.3~kpc)$ is one of 
the numerous dark clumps predicted by CDM scenarios, which was not 
destroyed by the galactic tidal field. It may be a disrupted cluster
as well, out of dynamical equilibrium. Here we assume that the NFW profile
is the signature of a halo made of cold particles. Physics beyond the
standard model could be supersymmetry. The lowest massive supersymmetric 
particle, i.e. the neutralino, is a natural 
candidate for CDM. If R-parity
is conserved the neutralino is stable, is its own antiparticle and has
a very small cross-section for annihilation. We assume that the halo of
Palomar 13 is made of neutralinos and calculate the flux in high energy
$\gamma$-rays due to their annihilation.
\vspace*{-0.1cm}
\section{Theoretical physics models, tools and benchmarks}
\vspace*{-0.1cm}
Theoretical physics beyond the standard model is reviewed by J. Ellis in 
this book. The SUSY benchmark models have been proposed to provide a common 
way of comparing the discovery potential of future accelerators [2]. These 
scenarios correspond to 13 configurations of the 5 mSUGRA parameters
with the trilinear coupling parameter $A_0$ set to 0. The models fulfill
the conditions imposed by LEP measurements, $g_\mu -2$ result and relic 
density constraint $0.1 < \Omega_\chi h^2 < 0.3$. We calculate the
$\gamma$ fluxes for the benchmark models BCGIL in the ``bulk'' region with 
our current MC simulation programs: DarkSUSY [3] and SUSPECT [4]. The 
simultaneous use of the SUSPECT and DarkSUSY package allows to perform the 
RGE evolution from the GUT scale to the EWSB scale.
\vspace*{-0.1cm}
\section{Predicted fluxes for Palomar 13}
\vspace*{-0.1cm}
The fluxes (in $10^{-12} cm^{-2} s^{-1}$) obtained by varying the 
threshold energy and integrated within $\Theta = 10^{-3}~sr$
are shown in the following figure.
\vspace*{-0.1cm}
\begin{figure}
\begin{center}
\includegraphics[width=.65\textwidth]{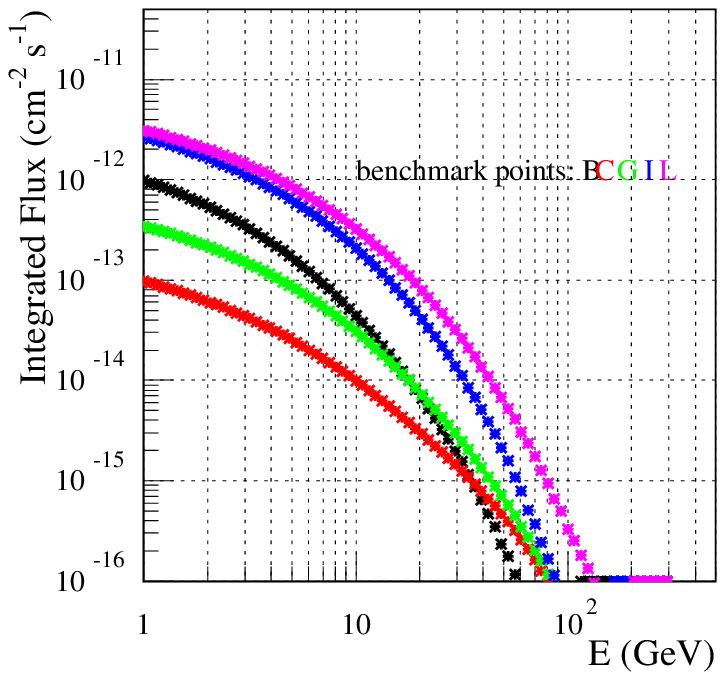}
\end{center}
\end{figure}
\vspace*{-2.0cm}
\section{An array of 16-20 Cerenkov telescopes at high altitude}
\vspace*{-0.1cm}
The main points are that
we need to work at low energy threshold and very low flux
$(10^{-12.5}~cm^{-2} s^{-1})$. These fluxes are not out of reach but 
ground-based ongoing instruments will have to be improved by one order 
of magnitude for that purpose. An array of 5 HESSes (each including four 
15-m class telescopes like in HESS), operating in adjacent areas at 5000m 
altitude, would reach a flux limit of 
$2.5 \times 10^{-13}~cm^{-2} s^{-1}$ at 25 GeV in 400 h. \bf With that 
flux limit most of the neutralino parameter space in the Galactic Center 
could be explored. \rm This is roughly the instrument needed to demonstrate or
disprove supersymmetry in astrophysics.

%

\end{document}